# Electron Attachment and Electron Ionization of Helium Droplets Containing Clusters of $C_{60}$ and Formic Acid

Masoomeh Mahmoodi-Darian,[1] Elias Jabbour Al Maalouf,[2,3] Samuel Zöttl,[2] Paul Scheier,[2] Olof Echt[2,4]

1    Department of Physics, Karaj Branch, Islamic Azad University, Karaj, Iran
2    Institut für Ionenphysik und Angewandte Physik, Universität Innsbruck, Technikerstr. 25, A-6020 Innsbruck, Austria
3    Faculté des Sciences IV, Laboratoire Energétique et Réactivité à l'Echelle Nanométrique (EREN), Université Libanaise, Haouch El-Omara, Zahlé, Lebanon
4    Department of Physics, University of New Hampshire, Durham NH 03824, USA

Correspondence to: Masoomeh Mahmoodi-Darian; e-mail: masoomeh2001@yahoo.co.uk,
    Olof Echt; e-mail: olof.echt@unh.edu

Running title: Helium droplets doped with $C_{60}$ and formic acid

**Abstract**

High-resolution mass spectra of helium droplets doped with $C_{60}$ and formic acid (FA) are ionized by electrons. Positive ion mass spectra reveal cluster ions $[(C_{60})_pFA_n]^+$ together with their hydrogenated and dehydrogenated counterparts. Also observed are ions containing one or more water (W) molecules. The abundance distributions of these ions reveal several interesting features: i) $[(C_{60})_pFA_n]^+$ ions are more abundant than hydrogenated $[(C_{60})_pFA_nH]^+$ ions even though the opposite is true in the absence of $C_{60}$ (i.e. if $p = 0$); ii) although $[C_{60}FA]^+$ is the most abundant ion containing a single $C_{60}$, multiple $C_{60}$ suppress the $[(C_{60})_pFA]^+$ signal; iii) an enhanced stability of $[(C_{60})_pW_1FA_5H]^+$ and $[(C_{60})_pW_2FA_6H]^+$ mirrors that of $[W_1FA_5H]^+$ and $[W_2FA_6H]^+$, respectively. On the other hand, the enhanced stability of $[C_{60}FA_6H]^+$ finds no parallel in the stability pattern of $[FA_nH]^+$ or $FA_n^+$. Negative ion mass spectra indicate a propensity for non-dissociated $[(C_{60})_pFA_n]^-$ anions if $p \geq 1$ which contrasts with the dominance of dehydrogenated $[FA_n-H]^-$ anions.

Keywords: Fullerene, formic acid, water, cluster, helium nanodroplets, electron attachment, ion-molecule reactions

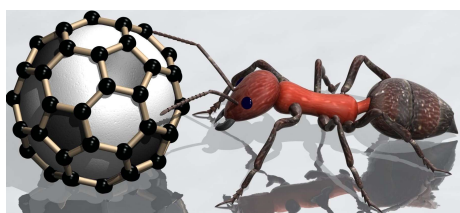
Graphical Abstract

## 1. Introduction

Formic acid (HCOOH, abbreviated FA) is the simplest of the carboxylic acids; it serves as a model system for the properties of larger, more complex molecules. It plays a role in Earth's atmospheric chemistry and is present in interstellar nuclear ice where its abundance exceeds that in interstellar gas by several orders of magnitude [1, 2]. It may play a key role in the formation of amino acids in the interstellar medium [3, 4].

The FA dimer is a prototype for double H-bonded cyclic complexes; it has been studied extensively [5-7]. The FA trimer has been characterized by infrared absorption in noble gas matrices [8]. It is chain-like, consistent with its polar character revealed in a molecular beam study utilizing an inhomogeneous electric field [9]. In the liquid bulk phase, X-ray and neutron diffraction indicate that FA molecules prefer to form short branched hydrogen bonded chains [10].

Charged clusters of FA have been produced in the gas-phase by photon, electron, proton and heavy ion-induced desorption from frozen FA films [2, 11-17]. In other work, neutral clusters have been formed by supersonic expansion of FA vapor, followed by photodissociation [18, 19], UV-ionization [20], and electron attachment [21]. Recently we adopted another approach by passing superfluid helium nanodroplets (HNDs) through a pickup cell filled with low-density FA vapor [22]. The doped droplets were subsequently bombarded with electrons resulting in positively and negatively charged FA cluster ions.

Surprisingly little research has been devoted to the interaction of formic acid with fullerenes. The combination shows promise for the development of direct formic acid fuel cells which are considered superior to direct methanol fuel cells thanks to their higher power density, efficiency and electromotive force [23, 24]. Palladium and lanthanum catalysts supported on modified $C_{60}$ have been shown to increase the electrocatalytic activity towards formic acid oxidation [25-27]. We are aware of only two reports of isolated FA-$C_{60}$ complexes. Petrie et al. have measured rate coefficients for proton transfer from $C_{60}^{2+}$ to FA in a selected-ion flow tube [28]. Oh et al. have investigated highly charged $C_{60}$ in a formic acid droplet by molecular dynamics simulations [29]. The strong electric field near the solvated ion might induce a conformational transition from trans-FA to cis-FA; large droplets would then exhibit star-like morphologies because of the high electric dipole moment of the cis conformer.

In the present work, we have doped HNDs with $C_{60}$ and FA; the doped droplets were subjected to electron ionization. Positive and negative ion mass spectra reveal the presence of $[(C_{60})_p(H_2O)_m FA_n]^{\pm}$ ions (where $p$, $m$, $n$ are small integers $\geq 0$) as well as their hydrogenated and dehydrogenated counterparts. In the presence of $C_{60}$, the preferred composition of cations changes from hydrogenated $[FA_nH]^+$ to undissociated $[(C_{60})_p FA_n]^+$; that of anions changes from dehydrogenated $[FA_n-H]^-$ to mostly undissociated $[(C_{60})_p FA_n]^-$. The suppression of the hydrogenation and dehydrogenation reactions is particularly strong for small values of $n$. Another interesting observation is the persistence of abundance anomalies (magic numbers) in complexes of FA with water (W): $[W_1 FA_5 H]^+$ and $[W_2 FA_6 H]^+$ form magic numbers with or without $C_{60}$. On the other hand, magic numbers in the distribution of $[C_{60} FA_n H]^+$ contrast with the smooth distribution of $[FA_n H]^+$. Lastly, the most abundant cation complexed with a single $C_{60}$ is $[C_{60} FA]^+$ but the $[(C_{60})_p FA]^+$ signal is strongly suppressed if $p > 1$.

## 2. Experiment

HNDs were produced by expanding helium (Linde, purity 99.9999 %) at a stagnation pressure of about 20 bar through a 5 μm nozzle, cooled by a closed-cycle cryostat (Sumitomo Heavy Industries LTD, model RDK-415D) to 9.35 K into vacuum. Droplets formed at these conditions contain an average of about $2 \times 10^6$ atoms [30]. The resulting supersonic beam was skimmed by a 0.8 mm conical skimmer, located 8 mm downstream from the nozzle. The skimmed beam traversed a differentially pumped pickup region into which a small amount of $C_{60}$ (SES Res., 99.99%) was vaporized from a crucible at 330 °C. The doped helium droplets passed through a second, 20-cm long differentially pumped pickup region into which FA (Sigma-Aldrich, 98%-100%) was introduced; the uncorrected pressure in that cell was $9 \times 10^{-6}$ mbar.

The doped helium droplets passed through a differentially pumped vacuum chamber (pressure $4 \times 10^{-8}$ mbar) where they were crossed with an electron beam of variable energy. Anions were formed at 21.5 eV, cations at 100 eV. Ions were accelerated into the extraction region of a commercial orthogonal time-of-flight mass spectrometer equipped with a reflectron (Tofwerk AG, model HTOF). The mass resolution was $m/\Delta m = 4500$ ($\Delta m$ = full-width-at-half-maximum); the base pressure in the mass spectrometer was $3 \times 10^{-7}$ mbar. The ions were detected by a micro-channel plate operated in single ion counting mode and recorded via a time-to-digital converter. Additional experimental details have been described elsewhere [31].

Mass spectra were evaluated by means of a custom-designed software [32]. The routine takes into account the isotope pattern of all ions that might contribute to a specific mass peak by fitting a simulated spectrum with defined contributions from specific atoms to the measured spectrum in order to retrieve the abundance of ions with a specific stoichiometry. The natural abundance of minor hydrogen and oxygen isotopes (0.0115 % for $^2$H, 0.038 % for $^{17}$O, 0.205 % for $^{18}$O) is small but the presence of $^{13}$C (1.07 %) leads to a multitude of mass peaks for each specific ion containing one or more $C_{60}$ (for an illustration, see refs. [32, 33]). The software also corrects for experimental artifacts such as background signal levels, the mass shift of the mass spectra, non-gaussian peak shapes, and mass drift over time.

## 3. Results and Discussion
### 3.1 Ionization of Doped Helium Droplets

Below, we will present positive and negative ion mass spectra that reveal $[(C_{60})_p W_m FA_n]^{\pm}$ ions where $p, m, n$ are small integers including 0. In addition, we observe hydrogenated cations and dehydrogenated cations or anions. We will graph the abundance of these ions versus $n$ because loss of FA molecules probably forms the energetically most accessible dissociation channel. The basic assumption in our discussion is that, at experimental conditions prevalent in the present study, local anomalies in the abundance distributions of cluster ions correlate with corresponding features in their dissociation energies $D_n$ (i.e. the energy required to adiabatically remove one FA molecule) [34, 35]. Cluster growth in a helium droplet by successive pick up of monomers is a statistical process which produces broad, featureless size distributions. Subsequent electron ionization introduces a large amount of excess energy which results in fragmentation and enrichment of relatively stable cluster ions.

For cations the process starts with the formation of $He^+$ in the droplet [36, 37]. The positive charge may jump by resonant charge exchange to an adjacent helium atom. This hopping process is terminated either by the formation of $He_2^+$ or by charge transfer to the dopant. In the latter case, about 15 eV (the difference between the ionization energies of helium, 24.6 eV, and the dopant) will be released [37, 38]. This excess energy exceeds the evaporation energy of bulk helium by a factor 24000.

For anions, the dominant process starts with the inelastic scattering of an incident electron off a helium atom (producing electronically excited $He^*$) and subsequent trapping of the thermalized electron in a bubble [37, 39]. The threshold energy for this channel equals the sum of the threshold energy to form $He^*$ (19.8 eV) and the energy needed for the incident electron to penetrate the surface of the droplet (1.2 eV). The slow electron can then attach to $He^*$ to form long-lived $He^{*-}$ which may migrate to the dopant. Electron transfer to the dopant will release the energy stored in $He^{*-}$ plus the electron affinity of the dopant.

The large excess energy leads to cluster ion ejection from the helium droplet and extensive fragmentation. Milliseconds elapse between ionization and mass analysis, providing ample time for the ions to cool by evaporation, enriching stable ions and depleting unstable ions. The relation between cluster abundance measured by mass spectrometry and (relative) dissociation energy is intricate [40, 41]. Qualitatively, however, it is clear that an abrupt decrease of the dissociation energy (i.e. $D_{n+1} \ll D_n$) will cause an enrichment of cluster ion $X_n^+$ at the expense of $X_{n+1}$. (The presence of alternative dissociation channels such as loss of $C_{60}$, a FA dimer, or electron detachment from an anion, will lead to a more complex scenario). There are several different scenarios, such as a single cluster size that is particularly stable (a "magic" cluster) or particularly unstable with respect to its neighbors, or closure of a solvation shell where $D_n$ drops in a stepwise fashion. Their signatures in mass spectra will be different but all of them are accompanied by abrupt decreases in the abundance relative to the envelope of the abundance distribution. For simplicity we will refer to cluster ions whose abundance is anomalously large relative to the next cluster size as "magic" or "particularly stable."

### 3.2 Positive Ions

A positive ion mass spectrum of helium droplets doped with $C_{60}$ and FA is displayed in Fig. 1. Panel a provides an overview (note the logarithmic ordinate). The most prominent mass peaks are due to bare fullerene clusters $(C_{60})_p^+$, $p \leq 12$; they are marked by asterisks. Panel b zooms into the low-mass region. Prominent mass peaks marked by full up triangles are due to $[FA_n H]^+$ ions (nominal mass $46 n + 1$ u). Also seen are mass peaks due to hydrogenated cluster ions complexed with a few water (W) molecules; they are marked by open up triangles and connected by dashed lines. In the low-mass region ($\leq 46$ u) our data are consistent with the NIST spectrum which shows the formyl cation $HCO^+$ at 100 %, $FA^+$ at 61 %, and $[FA-H]^+$ at 48 %; the abundance of other ions is below 20 % [42].

Other prominent mass peaks (not explicitly identified in Fig. 1b) are due to $He_n^+$, $C_n^+$, and $C_n^{2+}$. $C_n^+$ and $C_n^{2+}$ ions result from ionization of bare $C_{60}$. Presumably, a small amount of $C_{60}$ escapes from the pickup cell into the region of the ionizer; at an electron energy of 100 eV copious amounts of even-numbered $C_n^+$ above $n = 30$ and even as well as odd-numbered $C_n^+$ below $n = 30$ will be produced [33, 43, 44].

Fig. 1c reveals the presence of $C_{60}$ ions complexed with FA and water. Ions with the composition $[C_{60} FA_n]^+$ ($n = 0, 1, 2...$) that are isotopically pure (containing no $^{13}$C, $^2$H or $^{18}$O) appear at a nominal mass 720 + 46 $n$ u; they are marked by full circles. Another series of mass peaks appears at 720 + 18 $m$ + 46 $n$ u; they are due to $[C_{60}W_mFA_n]^+$ and marked by open circles. Dashed lines connect members of this series that share the same value of $n$. An expanded view of a positive ion mass spectrum is presented as Supplementary Material.

Each ion with a specific composition such as $[C_{60}W_mFA_n]^+$ produces a group of mass peaks because of the presence of $^{13}C$ (natural abundance 1.07 %). The characteristic isotope pattern of $C_{60}^+$ appears in the inset in Fig. 1c; ions containing several FA molecules will produce even broader multiplets that tend to obscure the appearance of hydrogenated ions $[C_{60}W_mFA_nH]^+$. Dehydrogenated ions are identified more readily as demonstrated in the inset of Fig. 1c: The satellite peak 1 u below $[C_{60}FA]^+$ (mass 766 u for isotopically pure ions) signals the presence of $[C_{60}FA-H]^+$.

The abundance of select cations is derived from the mass spectrum in Fig. 1 by a custom-designed software that takes into account all possible isotopologues, and contributions from impurities, background, and isotopologues of other ions [32]. Results are compiled in Fig. 2 for $[W_1FA_n]^+$, $[C_{60}FA_n]^+$, $[C_{60}W_1FA_n]^+$, and $[C_{60}W_2FA_n]^+$ (panels b, c, d, and e, respectively) together with their hydrogenated and dehydrogenated ions. Fig. 2a displays data for $FA_n^+$, $[FA_nH]^+$ and $[FA_n-H]^+$ that are taken from our previous report which featured better counting statistics and lower background because of the absence of $C_n^+$ fragments [22]. The abundance distributions of $[(C_{60})_pW_mFA_n]^+$ ($p$ = 1 through 4, $m$ = 0 or 1) and their hydrogenated counterparts are presented in Fig. 3 which employs a linear ordinate.

Figures 2 and 3 reveal several interesting features:

1. For ions not containing $C_{60}$ (Fig. 2a) the abundance of hydrogenated complexes is much higher than that of undissociated or dehydrogenated ions, but in the presence of $C_{60}$ hydrogenated ions are strongly suppressed if $n$ is small (less than approximately 6). This strong suppression also pertains to ions containing one or two water molecules (Fig. 2d, 2e). Also note that the abundance of $[C_{60}HCO]^+$ is an order of magnitude below that of $[C_{60}FA]^+$ whereas $HCO^+$ forms the most abundant peak in the NIST spectrum of FA [42].
2. The abundance distributions of $[FA_n]^+$ and $[FA_nH]^+$ are smooth but that of $[C_{60}FA_n]^+$ is slightly enhanced at $n$ = 4 and that of $[C_{60}FA_nH]^+$ is strongly enhanced at $n$ = 4, 5, 6. The enhancement comes at the expense of dehydrogenated ions $[C_{60}FA_n-H]^+$ but the sum of the three ion species (dashed line in Fig. 2c) shows a clear enhancement for $n$ = 4, 5, 6. For the purpose of the present discussion we consider $[C_{60}FA_4]^+$ and $[C_{60}FA_6H]^+$ to be magic clusters.
3. Series of hydrogenated ions containing a single water indicate enhanced abundance at $[(C_{60})_pW_1FA_5H]^+$, especially if $p$ = 0 or 1 (Fig. 2b, 2d).
4. Series of hydrogenated ions containing two water molecules indicate enhanced abundance at $[(C_{60})_pW_2FA_6H]^+$ for $p$ = 0 or 1 (Fig. 2b, 2e). A strong drop after $[(C_{60})_pW_2FA_4H]^+$ is noteworthy as well.
5. The most abundant ion containing one $C_{60}$ and at least one FA is $[C_{60}FA]^+$, but $[(C_{60})_pFA]^+$, $p$ > 1, is suppressed by more than an order of magnitude (Fig. 3).

We will discuss these observations in the order listed above. In the absence of any theoretical work on $C_{60}$-FA, the discussion will necessarily be qualitative and perhaps even speculative.

1. In previous work we have reported positive ion mass spectra of HNDs doped with $C_{60}$ and other molecules that, in the absence of $C_{60}$, tend to form hydrogenated cluster ions, including $H_2$ [33, 45, 46], $H_2O$ [47, 48], $NH_3$ [49], $CH_4$ [50, 51], $C_2H_4$ [52], $CH_3OH$, and $CH_5OH$ [53]; for a review see [54]. A comparison of our results for FA with other molecules that form hydrogen-bonded networks is most instructive. The electric dipole moments of $H_2O$, $NH_3$, and $C_3OH$ (1.85, 1.42, and 1.69 D, respectively [55]) are close to that of FA (1.5 D [29]). Moreover, these molecules share large proton affinities (7.16, 8.85, 7.82, and 7.75 eV for $H_2O$, $NH_3$, $CH_3OH$, and FA, respectively [56]) and similar ionization energies (12.62, 11.33, 10.84, and 10.07 eV, respectively [42]). For comparison, the proton affinity and ionization energy of $C_{60}$ are 8.75 eV [57] and 7.57 eV [42], respectively.

   Similar to FA, electron ionization of water, ammonia, and methanol clusters produces almost exclusively hydrogenated cluster ions [47-49, 58]. This can be attributed to an autoprotonation reaction. The reaction is endothermic but vertical ionization of, say, $(H_2O)_n$ results in ions that are energetically above the dissociation limit into $(H_2O)_{n-1}H^+$ + OH. One or more proton transfers take place in the water cluster within 0.1 ps, followed by ejection of OH and $H_2O$ [59]. In complexes containing $C_{60}$, however, the positive charge resides on the fullerene because of its low ionization energy, thus suppressing the autoprotonation reaction and loss of OH [47-49]. The re-appearance of hydrogenated cluster ions $[C_{60}X_nH]^+$ (X = $H_2O$, $NH_3$, FA) with increasing size $n$ is possibly linked to the decrease in the ionization energy of $X_n$ with increasing $n$. However, the abundances of $[C_{60}X_nH]^+$ and $[C_{60}X_n]^+$ become comparable at about $n$ = 6 for all X, in spite of substantial (2.5 eV) differences in their ionization energies. This suggests that some other mechanism plays a role, too. Perhaps the turnover occurs when the adduct cluster becomes three-dimensional? Theoretical work is needed to understand this transition at $n \approx 6$.

2. The enhanced ion abundance of $[C_{60}FA_4]^+$ and $[C_{60}FA_nH]^+$ for $n = 4, 5, 6$ contrasts with the smooth distributions of $[FA_n]^+$, $[FA_nH]^+$, and $[C_{60}FA_n]^+$. Feng and Lifshitz had mentioned an enhanced abundance of $[FA_nH]^+$ at $n = 6$ for clusters formed in a pressure-variable ion source [60]. However, close inspection of their data suggests that $n = 6$ accidentally happened to be the maximum of the abundance distribution; the maximum shifted to $n = 5$ or 4 when the experimental conditions were changed. Heinbuch et al. reported a weak enhancement of the $[FA_5H]^+$ abundance in mass spectra of FA clusters formed in a supersonic expansion followed by single photon ionization [20].

   Thus, experimental data suggest that the dissociation energies of $[FA_nH]^+$ vary rather smoothly with $n$. However, ab-initio calculations indicate an interesting structural transition from open chains with free OH groups at both ends for $n \leq 5$ to a structure where one or both ends are terminated by cyclic dimers for $n > 5$ [19, 61]. Infrared photodissociation spectroscopy and the observation of unimolecular $FA_2$ loss from cluster ions larger than $n = 5$ support these conclusions [19, 60]. It is not obvious, however, how this structural transition and competition between monomer and dimer loss will stabilize specific cluster sizes in the presence of $C_{60}$. Furthermore, we note that the enhancement of $[(C_{60})_pFA_4]^+$ and $[(C_{60})_pFA_4H]^+$ persists when multiple $C_{60}$ are present (see Fig. 3), but $[(C_{60})_pFA_6H]^+$ is no longer enhanced for $p > 1$.

3. Several groups have reported ion abundance distributions of $[W_1FA_nH]^+$, employing liquid ionization mass spectrometry [62], a variable pressure cluster source [60], single-photon and electron ionization of a supersonic cluster beam [18, 20], and ion desorption from frozen films of FA exposed to energetic particles and photons [15, 17]. Universally, the $[W_1FA_5H]^+$ was found to be strongly enhanced in agreement with our data (Fig. 2b). Theory indicates that $H_3O^+$ resides at the center of a cyclic FA pentamer [18, 63]. One may conjecture that the structure of the magic, cyclic $[W_1FA_5H]^+$ remains more or less unchanged in the presence of one or more $C_{60}$. However, reality is probably not as simple because the proton affinity of $C_{60}$ exceeds that of FA by 1.0 eV and that of water by 1.6 eV. Also note that the core of $[W_1FA_nH]^+$ changes from a hydrogenated FA ion for $n \leq 3$ to a hydrogenated water ion for larger clusters [18, 63, 64].

   It is worth mentioning that the relative abundance of ions containing one or more $H_2O$ molecules is unusually large. This was already pointed out in our previous electron ionization study of formic acid clusters where a possible water contamination in the background gas or the helium droplets was deemed unlikely but could not be ruled out completely [22]. Now, however, we observe that the abundance of $[C_{60}H_2O]^+$ is much weaker than that of $C_{60}^+$, and that of $[C_{60}(H_2O)_2]^+$ is weaker by another two orders of magnitude, see Fig. 1c. This suggests that water-FA complexes arise from an intracluster ion-molecule reaction rather than a significant water contamination in the HNDs. Bernstein and coworkers have suggested that $[W_1FA_nH]^+$ arises from $[FA_{n+1}H]^+$ via loss of a CO molecule [20]. Similarly, we have attributed the large abundance of hydrgoenated methanol cluster ions complexed with several water molecules to consecutive ion-molecule reactions that lead to multiple loss of $C_2H_6O$ [58]. However, methanol-water complexes were absent below the methanol heptamer whereas complexes of FA with water appear for any number of FA molecules. Furthermore, methanol-water complexes were relatively weak in the presence of $C_{60}$ [53], contrary to our present results for FA-water.

4. $[W_2FA_nH]^+$ ions can be identified in mass spectra reported by Feng and Lifshitz who used a pressure-variable ion source [60] and Heinbuch et al. who studied single-photon ionization of clusters in a supersonic beam [20]. Their data feature a magic $[W_2FA_6H]^+$ and a less pronounced enhancement for $[W_2FA_4H]^+$, in agreement with our findings (Fig. 2b). Spectra obtained by Andrade et al. by particle and radiation bombardment of cryogenic FA films show similar features although at lower resolution and poorer statistics [15]. Goken and Castleman have formed hydrogenated water-FA cluster ions in a fast-flow reactor [65]. However, their focus was on large ($n \approx 21$) water clusters complexed with a few FA molecules; the species identified as magic in our work are not present in their mass spectra.

   We are not aware of any theoretical work pertaining to $[W_2FA_nH]^+$ ions. Guessing the structures of the magic $[W_2FA_4H]^+$ and $[W_2FA_6H]^+$ based on structures computed for $[W_1FA_nH]^+$, $n \leq 8$ [63], would be unwise. We emphasize, though, that the "magic character" of $[C_{60}W_2FA_6H]^+$ is dramatic; its abundance exceeds that of $[C_{60}W_2FA_5H]^+$ by an order of magnitude.

5. $[C_{60}FA]^+$ is the most abundant ion in the $[C_{60}FA_n]^+$ series (Fig. 3). However, upon adding one or more $C_{60}$, the abundance of $[(C_{60})_pFA]^+$ relative to that of $[(C_{60})_p FA_2]^+$ drops from $>1$ ($p = 1$) to $< 0.1$ ($p > 1$). A similar phenomenon was observed for mixed clusters of $NH_3$ and $C_{60}$, where the abundance of $[(C_{60})_pNH_3]^+$ relative to that of $[(C_{60})_p(NH_3)_2]^+$ dropped from about 0.3 ($p = 1$) to $< 0.02$ ($p = 2, 3$) [49].

Another system which shows the absence of some small cluster ions are complexes of $C_{60}$ and gold or copper [66]. Complexes of $C_{60}$ and water, however, do not show a related behavior [48, 49].

Ions observed in the present study are products of dissociative ionization of cluster ensembles that grow statistically in HNDs. In most cases, observed ions $X_n^+$ are the products of monomer evaporation from $X_{n+1}^+$ which, in turn, may be the product of monomer evaporation from $X_{n+2}^+$, and so on. Logically, the near-absence of a specific size $n$ may be explained by i) its extreme instability, causing it to rapidly dissociate further into $X_{n-1}^+$, or ii) dissociative reactions that by-pass $X_n^+$ altogether, for example if $X_{n+1}^+$ dissociates by dimer emission rather than monomer emission (but $X_{n+2}^+$ dissociates by monomer rather than dimer emission). For binary systems there is a third possibility, for example iii) dissociation of $[(C_{60})_2X_2]^+$ into $C_{60}^+ + [C_{60}X_2]$ or into $X^+ + [(C_{60})_2X]$. This third route is a realistic option only if X is a metal atom with an ionization energy below that of $C_{60}$, or with a $X_2$ bond strength exceeding that of the $C_{60}$ dimer. For the $C_{60}$-FA system route (ii) appears more likely because the neutral FA dimer is strongly bound (0.67 eV) [5]; at room temperature and normal pressure, 95% of formic acid vapor consists of the dimer [60]. Unimolecular dissociation of $[FA_nH]^+$ proceeds by monomer loss for $n \leq 6$ but loss of dimers for $n = 7, 8$ [60]. Calculations are needed to see if dimer evaporation can explain the striking depletion of $[(C_{60})_pFA]^+$ for $p > 1$. Ab-initio calculations for $[C_{60}W_n]^+$ showed that the water cluster was weakly bound to $C_{60}^+$, and that desorption of the complete water cluster energetically competes with evaporation of single $H_2O$ for $n = 3, 4$, and 6, in agreement with experimental data [48]. Unfortunately, our current experimental setup does not allow to study unimolecular dissociation of $[(C_{60})_pFA_n]^+$.

So far we have attempted to attribute the near-absence of $[(C_{60})_pFA]^+$, $p > 1$, to dimer evaporation from $[(C_{60})_pFA_2]^+$. We have not addressed why ions containing a single $C_{60}$ behave differently because, in the absence of theoretical work, we do not have a compelling explanation for the difference.

### 3.3 Negative Ions

A mass spectrum of negative ions is displayed in Fig. 4; an expanded view is presented as Supplementary Material. Prominent ions in the low-mass region (panel a) are $O^-$ and $OH^-$. Two series of cluster ions, $[FA_n-H]^-$ and the less abundant $[FA_n]^-$ are marked by down triangles and circles, respectively. Also seen is a weak series of $[W_1FA_n-H]^-$ (open circles). Panels b through d reveal various anions containing one $C_{60}$. Each section spans 100 u, starting just below $[C_{60}FA_n]^-$ ($n = 0, 2, 4$, respectively) and ending just above $[C_{60}FA_{n+2}]^-$. Mass peaks due to isotopically pure $[C_{60}FA_n]^-$ and $[C_{60}W_mFA_n]^-$ are marked by full and open circles, respectively. Logarithmic ordinates were chosen in panels a and b, but linear ordinates in panels c and d.

The abundance distributions of $FA_n^-$ and $[FA_n-H]^-$ are presented in Fig. 5a. Figs. 5b, 5c, and 5d display the distributions of $[(C_{60})_pFA_n]^-$ and $[(C_{60})_pFA_n-H]^-$ for $p = 1, 2, 3$, respectively. Distributions of $[C_{60}W_1FA_n]^-$ and $[C_{60}W_1FA_n-H]^-$ are shown in panel e. Interesting features in these data are as follows:

1. For bare formic acid cluster ions (panel a), dehydrogenated anions $[FA_n-H]^-$ are much more abundant than their non-dissociated counterparts $FA_n^-$. The abundance of $FA^-$ and $FA_2^-$ is close to zero; it rises abruptly above the dimer. As discussed in our previous work, the adiabatic electron affinity (AEA) of FA is negative, i.e. $FA^-$ is metastable at best. An ab-initio calculation with the 6-311++G** basis set by Ziemczonek and Wroblewski yielded an AEA of -1.27 eV for FA [67]. Valadbeigi and Farrokhpour obtained similar values using CBS-Q, G4MP2, W1BD, and G2MP2 methods [68]. The calculated AEA increases dramatically from -1.16 eV for the dimer to -0.09 eV for the trimer [67]. Thus, the vertical detachment energy of $FA_3^-$ could be substantial; a cold trimer ion would be long-lived. The abundance of $[FA_n-H]^-$ peaks at $n = 3$ but we do not assign enhanced stability to this ion; in our previous work the abundance showed a broader maximum between $n = 3$ and 5 [22].

2. In the presence of one or more $C_{60}$, non-dissociated $FA_n^-$ becomes more abundant than $[FA_n-H]^-$. $C_{60}$ has a positive electron affinity of 2.68 eV [42]; the AEA of the dimer is probably near 3.5 eV [69-71]. Therefore the excess electron in $[(C_{60})_pFA_n]^-$ will reside on the fullerene and the anionic complex might be best described as a FA cluster weakly bound to $(C_{60})_p^-$. Interestingly, the abundance of $[(C_{60})_pFA_n-H]^-$ approaches that of $[(C_{60})_pFA_n]^-$ at $n \approx 6$, the same size where the abundances of $[(C_{60})_pFA_nH]^+$ and $[(C_{60})_pFA_n]^+$ become about equal. We cannot explain why the abundances of different homologous ion series should converge at about the same size, independent of the charge state.

3. Considering homologous ion series $[(C_{60})_pFA_n]^-$ versus $n$, the first member ($n = 1$) has maximum abundance if $p = 1$, but it is strongly suppressed if $p > 1$. As result, $[(C_{60})_pFA_2]^-$ ions form local maxima if $p > 1$. Another local maximum appears at $n = 4$. These features are akin to those observed for cations (Fig. 3). However, no feature analogous to the enhancement of hydrogenated cations containing one water molecule at $n = 5$ is seen for anions.

4. The abundance of $[W_1FA_n-H]^-$ is weak relative to that of $[FA_n]^-$ or $[FA_n-H]^-$ whereas in the presence of $C_{60}$ the abundances of anions with or without water are comparable (see Fig. 4). If it is true that $[W_1FA_n-H]^-$ results from an intracluster reaction that is followed by loss of HCO from $FA_n^-$ (or $HCO^+$ from neutral $FA_n$) [72], then the enhanced abundance of $[(C_{60})_pW_1FA_n]^-$ might indicate catalytic activity of $C_{60}$ in this reaction.

**Conclusion**

We have presented positive and negative ion mass spectra of helium nanodroplets doped with $C_{60}$ and FA. There are marked differences in the preferred stoichiometry of ions upon the addition of one or more $C_{60}$. The dominant ion series have the composition $[(C_{60})_pFA_n]^\pm$ if $p \geq 1$ whereas, in the absence of $C_{60}$, the most abundant ion series are hydrogenated cations $[FA_nH]^+$ and dehydrogenated anions $[FA_n-H]^-$. We have tentatively attributed these differences to charge localization on $C_{60}$ whose ionization energy is lower than that of FA clusters whereas its electron affinity is higher. Also noteworthy are similarities in the stability patterns of ions containing one or two water molecules, $[(C_{60})_pW_mFA_nH]^+$ ($p \geq 1$) with stability patterns for $p = 0$. On the other hand, the enhanced stability of $[C_{60}FA_4H]^+$ and $[C_{60}FA_6H]^+$ is not mirrored in an enhanced stability of $[FA_4H]^+$ or $[FA_6H]^+$, respectively. In the absence of any theoretical data our discussion remained necessarily qualitative. Hopefully, the present results will stimulate theoretical work that elucidates the properties of neutral and charged complexes of $C_{60}$ with FA.

Supplementary Material
See Supplementary Material for expanded views of mass spectra.

Funding
This work was supported by the Austrian Science Fund, FWF (Projects P23657, I4130, and P31149), and the European Commission (ELEvaTE H2020 Twinning Project, Project No. 692335)

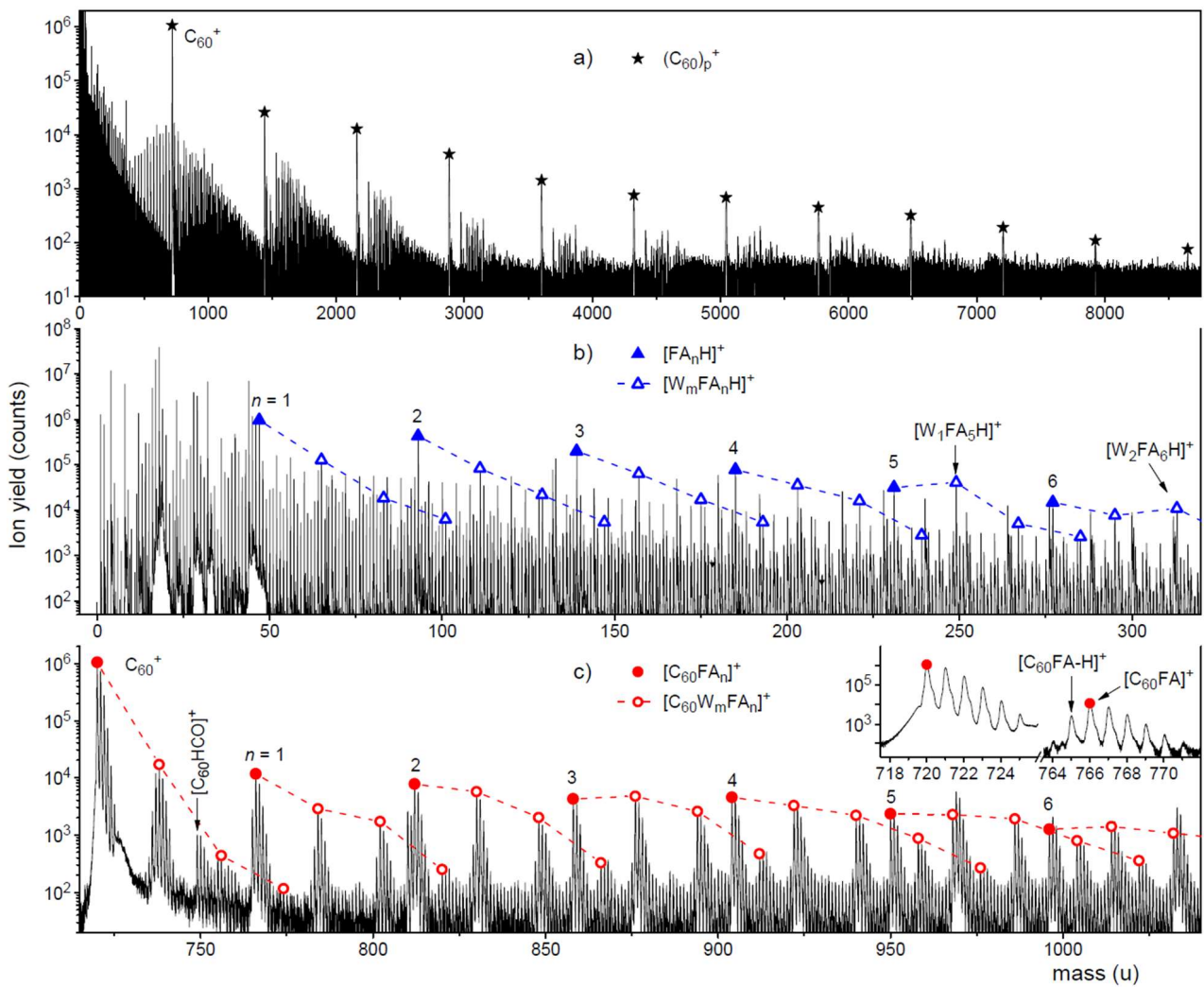

**Fig. 1**: Positive ion mass spectrum of helium nanodroplets (HNDs) doped with $C_{60}$ and formic acid (FA, mass 46 u). The most prominent ion series in panel a is due to $(C_{60})_p^+$, $p \leq 12$. Panel b zooms into the low-mass region where hydrogenated FA cluster ions (full up triangles) dominate. Complexes of $[FA_nH]^+$ with one or more water (W) molecules are flagged by open up triangles. The most prominent mass peaks above 720 u (isotopically pure $^{12}C_{60}^+$) are due to $[C_{60}W_mFA_n]^+$, $m = 0, 1, 2,…$ (panel c, all symbols refer to isotopically pure ions). The inset in panel c reveals the presence of dehydrogenated ions, $[C_{60}FA-H]^+$.

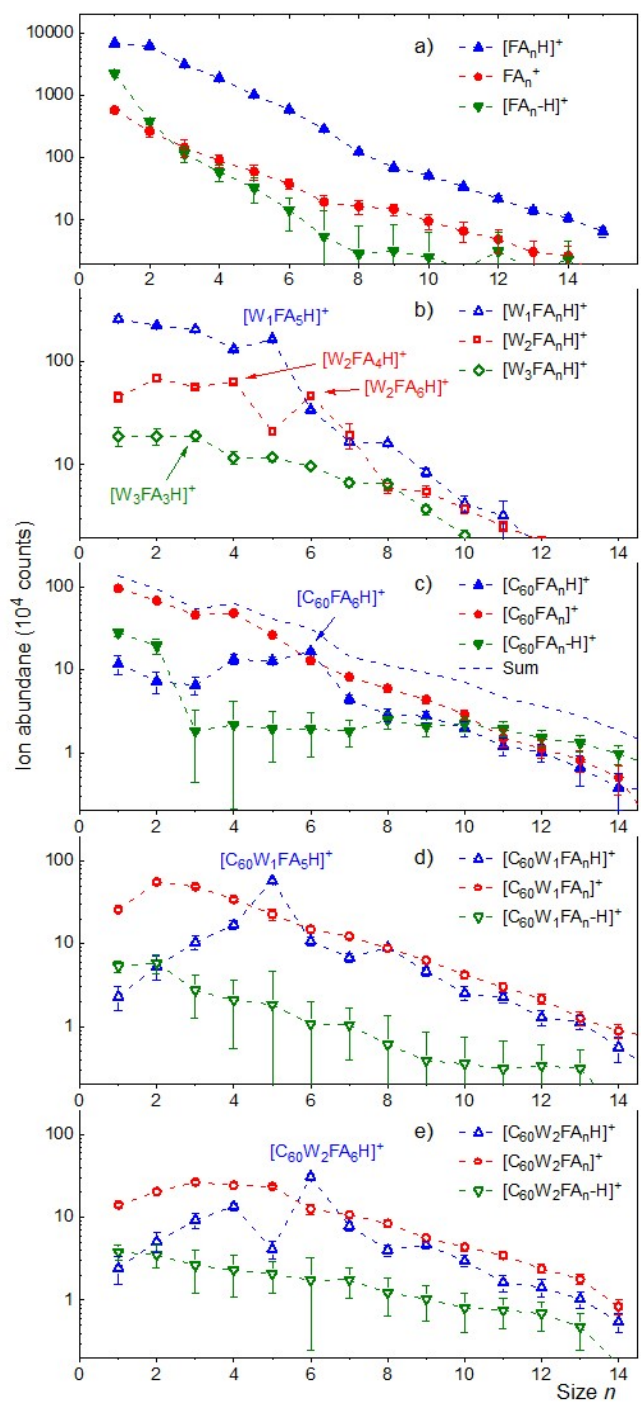

**Fig. 2**: Abundance distributions of FA cluster cations extracted from the mass spectrum in Fig. 1 that contain zero or one $C_{60}$ plus zero, one, or two water molecules.

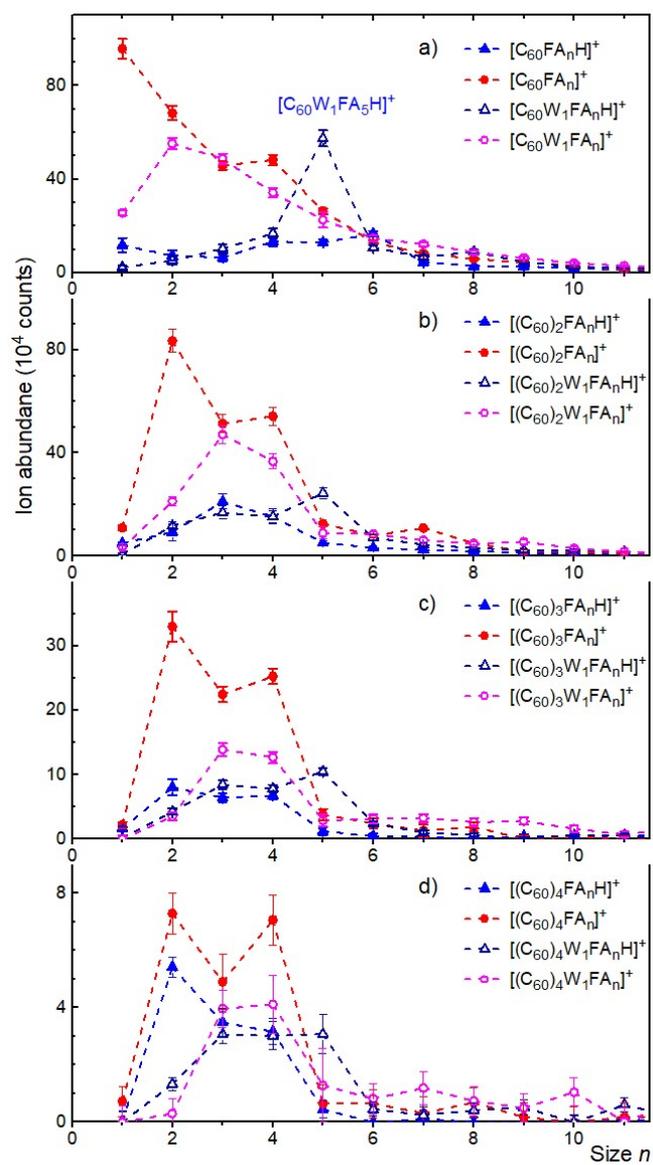

**Fig. 3**: Abundance distributions of FA cluster cations that contain one to four $C_{60}$ plus zero or one water molecule.

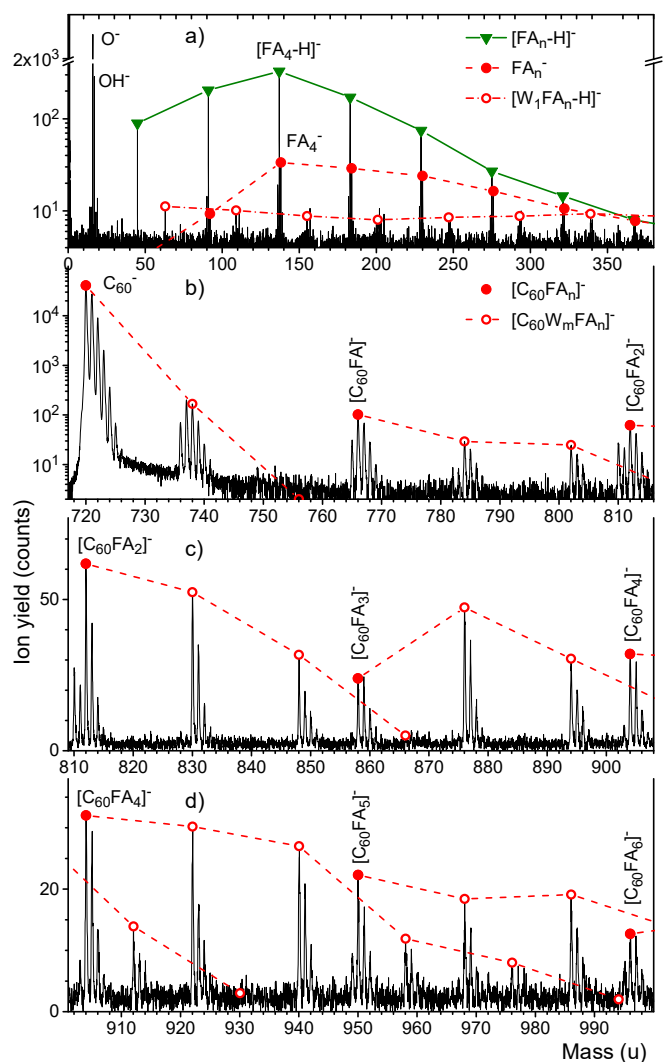

**Fig. 4**: Sections of a negative ion mass spectrum of HNDs doped with $C_{60}$ and FA. Dehydrogenated $[FA_n-H]^-$ ions dominate at low mass (panel a) whereas $[C_{60}FA_n]^-$ and $[C_{60}W_mFA_n]^-$ (full and open dots, respectively) dominate above the mass of $C_{60}^-$ (panels b through d).

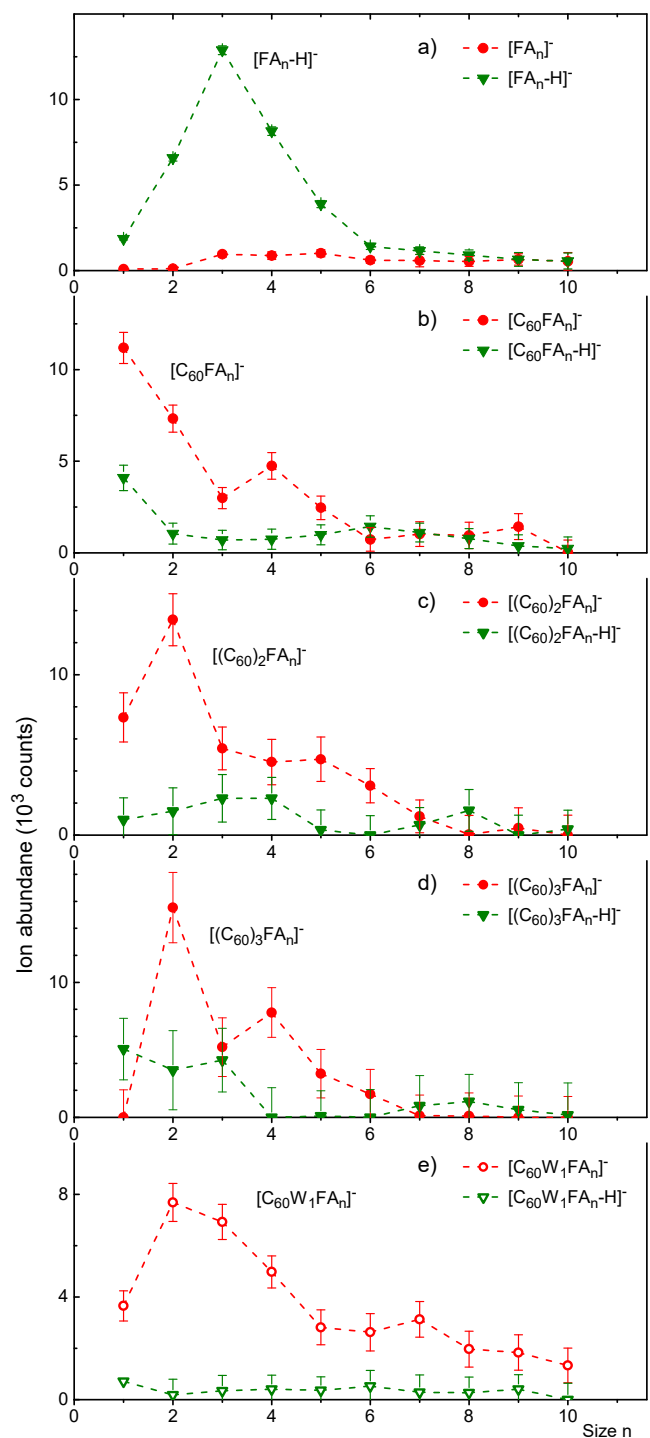

**Fig. 5:** Abundance distributions of FA cluster anions that contain up to three $C_{60}$ plus zero or one water molecule.